\documentclass[%
reprint,
superscriptaddress,
 amsmath,amssymb,
 aps,
]{revtex4-2}

\usepackage{graphicx}
\usepackage{dcolumn}
\usepackage{bm}
\usepackage{xcolor}
\usepackage{csquotes}

\begin{document}

\preprint{APS/123-QED}

\title{Measurement of zero-frequency fluctuations generated by coupling between Alfvén modes in the JET tokamak}



\author{J. Ruiz Ruiz}
\email[]{juan.ruiz@physics.ox.ac.uk}
\affiliation{Rudolf Peierls Centre for Theoretical Physics, University of Oxford, OX1 3NP, UK}

\author{J. Garcia}%
    \affiliation{CEA, IRFM, F-13108 Saint-Paul-lez-Durance, France}

\author{M. Barnes}
 \affiliation{Rudolf Peierls Centre for Theoretical Physics, University of Oxford, OX1 3NP, UK}
 \affiliation{University College, Oxford OX1 4BH, UK}
 
\author{M. Dreval}
\affiliation{National Science Center Kharkiv Institute of Physics and Technology, 1 Akademichna Str., Kharkiv 61108, Ukraine}

\author{C. Giroud}
 \affiliation{UKAEA (United Kingdom Atomic Energy Authority), Culham Campus, Abingdon, Oxfordshire, OX14 3DB, UK}

\author{V. H. Hall-Chen}%
 \affiliation{Institute of High Performance Computing, A*STAR, Singapore 138632, Singapore}

\author{M. R. Hardman}
 \affiliation{Tokamak Energy Ltd, 173 Brook Drive, Milton Park, Abingdon, OX14 4SD, United Kingdom}

\author{J. C. Hillesheim}
 \affiliation{Commonwealth Fusion Systems, Devens, MA, USA}

\author{Y. Kazakov}%
 \affiliation{Laboratory for Plasma Physics, ERM/KMS, TEC Partner, 1000 Brussels, Belgium}

\author{S. Mazzi}%
 \affiliation{CEA, IRFM, F-13108 Saint-Paul-lez-Durance, France}

\author{B. S. Patel}
 \affiliation{UKAEA (United Kingdom Atomic Energy Authority), Culham Campus, Abingdon, Oxfordshire, OX14 3DB, UK}

\author{F. I. Parra}%
 \affiliation{Princeton Plasma Physics Laboratory, Princeton, New Jersey 08543, USA}

\author{A. A. Schekochihin}%
 \affiliation{Rudolf Peierls Centre for Theoretical Physics, University of Oxford, OX1 3NP, UK}
 \affiliation{Merton College, Oxford OX1 4JD, UK}

\author{Z. Stancar}%
 \affiliation{UKAEA (United Kingdom Atomic Energy Authority), Culham Campus, Abingdon, Oxfordshire, OX14 3DB, UK}

\author{the JET contributors}
 \affiliation{See the author list of “Overview of T and D-T results in JET with ITER-like wall” by C. F. Maggi et al., 2024 Nucl. Fusion 64 112012.}

\author{the EUROfusion Tokamak Exploitation Team}
 \affiliation{See the author list of “Overview of the EUROfusion Tokamak Exploitation programme in support of ITER and DEMO” by E. Joffrin Nuclear Fusion 2024 10.1088/1741-4326/ad2be4}


\date{\today}

\begin{abstract}

We report the first experimental detection of a zero-frequency fluctuation that is pumped by an Alfvén mode in a magnetically confined plasma. Core-localized Alfvén modes of frequency inside the toroidicity-induced gap (and its harmonics) exhibit three-wave coupling interactions with a zero-frequency fluctuation. The observation of the zero-frequency fluctuation is consistent with theoretical and numerical predictions of zonal modes pumped by Alfvén modes, and is correlated with an increase in the deep core ion temperature, temperature gradient, confinement factor $H_{89,P}$, and a reduction in the main ion heat diffusivity. Despite the energetic particle transport induced by the Alfvén eigenmodes, the generation of a zero-frequency fluctuation that can suppress the turbulence leads to an overall improvement of confinement.




\end{abstract}


\maketitle


Achieving net energy gain from magnetically confined thermonuclear fusion is primarily hindered by the turbulent transport of heat and particles across the confining magnetic field. Turbulent transport determines the background plasma profiles, and, as a result, the overall fusion power-balance in a reactor. In a burning plasma, fusion reactions between deuterium and tritium (DT) will generate supra-thermal alpha particles of energy $E=3.5$ MeV, more than two orders of magnitude higher than the temperature of the thermal plasma, typically $T \approx 15-30 $ keV. Fusion-born alpha particles are expected to have characteristic velocities larger than the Alfvén speed $v_A = B/(\mu_0 \sum_i m_i n_i)^{1/2}$, where $B$ is the background magnetic field, $\mu_0$ is the vacuum magnetic permeability and the sum is taken over the thermal ion species of mass $m_i$ and density $n_i$. Such a super-Alfvénic population is capable of triggering magnetohydrodynamic (MHD) Alfvén instabilities, which have been predicted \cite{rosenbluth_prl_1975} and observed in a wide range of magnetic-confinement devices \cite{heidbrink_nf_1994, fasoli_nf_2007, vanzeeland_pop_2011, sharapov_nf_2013, gorelenkov_nf_2014, cecconello_ppcf_2015}. For many years, the presence of fast-ion-driven Alfvén instabilities has been thought to be detrimental to the overall plasma confinement, since they are detrimental for the confinement of the energetic particles themselves \cite{heidbrink_pop_2008, todo_revmodphys_2018, heidbrink_pop_2020}. Recently, however, numerical simulations of the turbulence have shown that fast-ion-driven Alfvénic activity can have a positive effect on the confinement of the \emph{thermal} plasma \cite{biancalani_ppcf_2021, mazzi_natphys_2022, Garcia_2022, biancalani_jpp_2023, citrin_ppcf_2023}. The current leading hypothesis explaining this effect is the pumping by the Alfvén eigenmode of a zonal-flow component, which could interact with, and suppress, the turbulence. This seems to be in agreement with analytical predictions of zonal-flow excitation by Alfvén eigenmodes \cite{chen_prl_2012, qiu_pop_2016, qiu_nf_2017, qiu_revmodphys_2023} and with numerical simulations \cite{spong_pop_1994, todo_nf_2010, todo_nf_2012, zhang_pst_2013, biancalani_iaea_2016}. Zonal modes are zero-frequency perturbations that are constant within a magnetic-flux surface. In the traditional zonal-flow paradigm, a zonal flow generated by the turbulence fluctuations can produce stabilization \cite{lin_science_1998, rosenbluth_prl_1998, rogers_prl_2000, chen_pop_2000, diamond_ppcf_2005}. The zonal-flow generation by Alfvén eigenmodes, however, has so far lacked direct experimental confirmation. In this article, we report the first experimental detection of a zero-frequency fluctuation that is pumped by Alfvén eigenmodes.

\begin{figure}
\begin{center}
	\includegraphics[height=9cm]{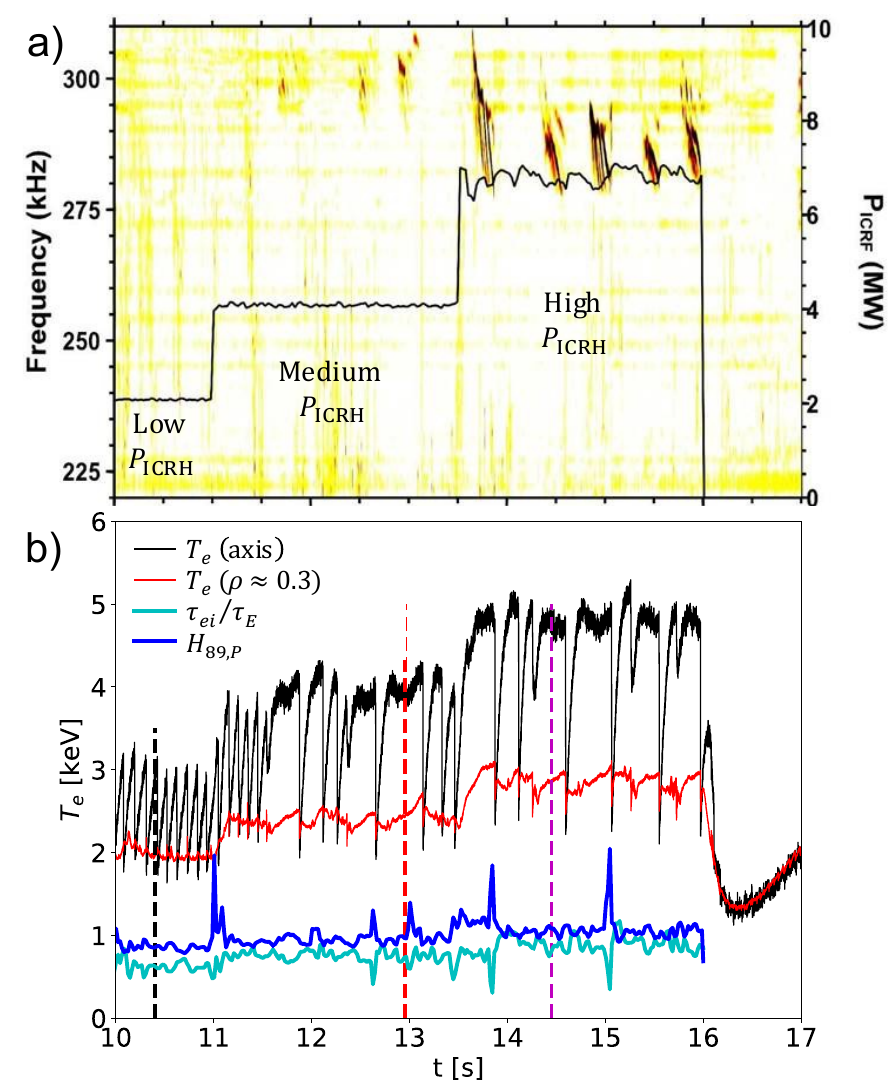}
\end{center}
\caption{(a) Line-integrated, density fluctuation spectrogram from the FIR interferometer in JET, along with the total ICRH power $P_\text{ICRH}$. (b) Electron temperature from the ECE diagnostic on axis (black) and at $\rho\approx 0.3$ (red). Also shown are $\tau_{ei}/\tau_E$ (cyan), the electron-ion collision time divided by the energy confinement time (TRANSP), and, the L-mode confinement factor $H_{89,P}$ (blue). Vertical dashed lines indicate times for which DBS measurements are shown in figure \ref{dbs_med}.}   
\label{magnetics}
\end{figure}

The Joint European Torus (JET \cite{rebut_fustech_1987}) is the magnetic-confinement facility that has come the closest to approaching the physics regimes of a burning plasma \citep{mailloux_nf_2022, kiptily_prl_2023, maggi_nf_2024, garcia_natcomm_2024}. The JET discharge 97090 was an L-mode characterized by a plasma current of $I_p = 2.4$ MA, toroidal magnetic field $B_0 = 3.2$ T, line-integrated density $\int n_e \text{d}l = 7.5 \times 10^{19}$ m$^{-2}$, Greenwald fraction of $f_\text{GW} = 0.36$ \cite{greenwald_ppcf_2002} and $q_{95} = 3.75$, all of which remained constant within uncertainties throughout the discharge (the central electron number density increased from $\approx 3.5$ to $4 \times 10^{19}$ [m$^{-3}$] as the discharge progressed). The plasma was ion-cyclotron-resonance heated (ICRH) using the three-ion scheme \cite{kazakov_nf_2015, kazakov_pop_2015, kazakov_natphys_2017} (main ions were composed of $80\%$ hydrogen and $20\%$ deuterium). A first Ohmic phase was followed by phases of increasing ICRH power of $P_\text{ICRH} = $ 2 MW (low), 4 MW (medium) and 7 MW (high, see black curve in figure \ref{magnetics}.(a)). A trace population of $^3$He ions with density $n_{^3\text{He}}/n_e \approx 0.2-0.3 \% $ absorbed $ \gtrsim 90 \%$ of the ICRH power and was accelerated to energies in the deep core of $E_{^3\text{He}} \approx 1.4, 4$ and $5$ MeV respectively in the low-, medium-, and high-$P_\text{ICRH}$ phases. These values are predicted by TRANSP \cite{transp} and TORIC \cite{brambilla_ppcf_1999}, and are consistent with $\gamma$-ray measurements in similar discharges \cite{nocente_nf_2020}. In the absence of tritium, MeV-range $^3$He can mimic the effect of fusion-born alpha particles in a burning plasma (although the anisotropic ICRH fast-ion distribution function is qualitatively different from the isotropic alpha-particle fusion source, and can result in differences in the Alfvénic-instability drive \cite{wong_physlettA_1999}). The fast $^3$He ions deposited $\approx 90 \%$ of their energy on the thermal electrons by collisionally slowing down. The remainder of the ICRH power is predicted to be absorbed directly by electrons, with less than $ 1\%$ of the power predicted to be absorbed directly by the thermal ions. Thus, $ \gtrsim 90 \%$ of the heating power was predicted to be ultimately deposited on the electrons. As we will see, this is at odds with the increase in the deep-core ion temperature observed in this plasma.




\begin{figure}
\begin{center}
        \includegraphics[height=10cm]{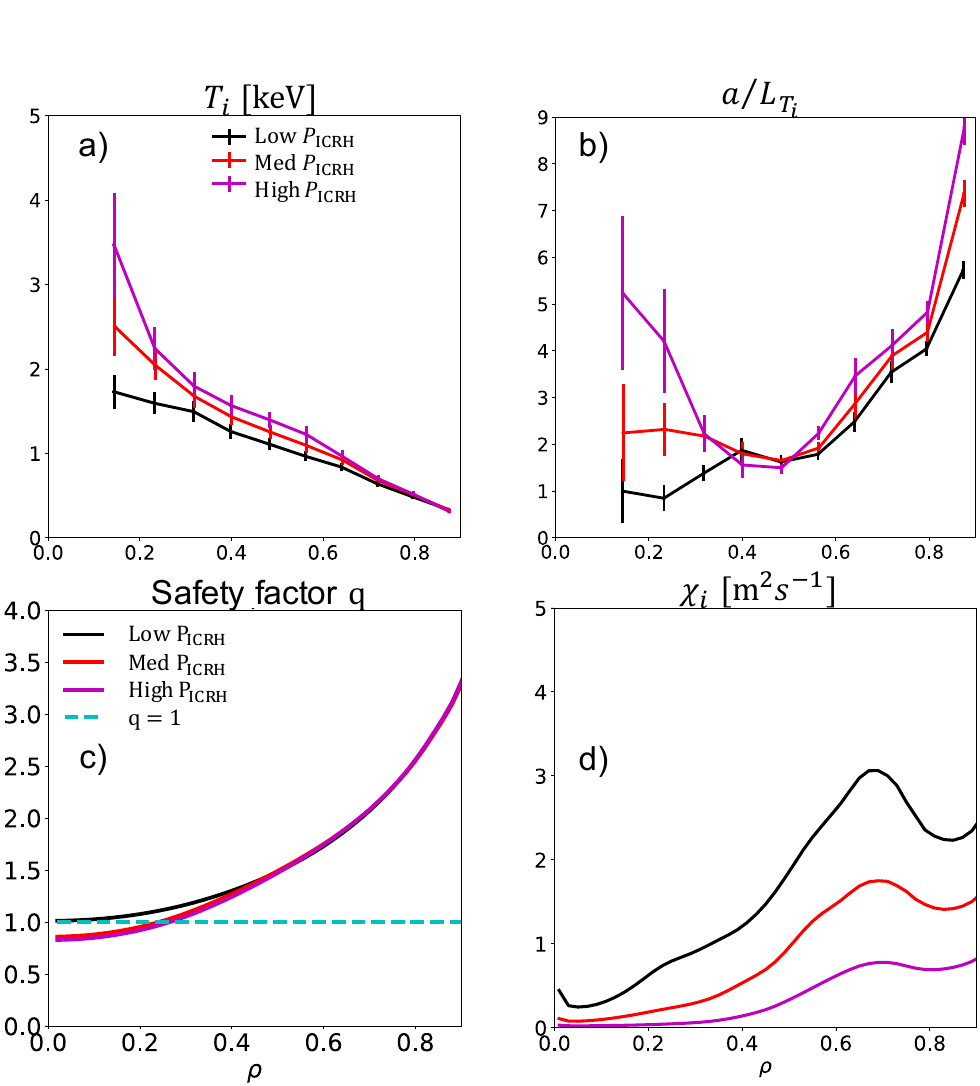}
\end{center}
\caption{(a) Profiles of the main-ion temperature $T_i$ at low, medium, and high $P_\text{ICRH}$. (b) Normalized temperature-gradient scale length $a/L_{Ti}$. Both $T_i$ and $a/L_{Ti}$ increase inside $\rho\approx 0.4$ with increasing $P_\text{ICRH}$, despite negligible main-ion heating. (c) Safety factor profile. (d) Main-ion heat diffusivity computed by TRANSP \cite{transp}. }   
\label{teti_time}
\end{figure}

The plasma exhibits internal sawtooth oscillations \cite{goeler_prl_1974} with an inversion radius at $\rho \approx 0.25-0.3$ (consistent with the safety factor $q=1$, figure \eqref{teti_time}.(c)), where $\rho$ is the magnetic-flux-surface label defined at the square root of the normalized toroidal magnetic flux ($\rho=0$ corresponds to the magnetic axis, while $\rho=1$ is the location of the edge separatrix). The sawtooth period increases with $P_\text{ICRH}$, as shown by the electron temperature $T_e$ in figure \ref{magnetics}.(b) (from the electron-cyclotron-emission diagnostic, ECE), and consistent with the stabilization of sawteeth by fast ions \cite{campbell_prl_1988}. This gives rise to the so-called monster sawteeth during the medium- and high-$P_\text{ICRH}$ phases \citep{bernabei_prl_2000}. Following a sawtooth event, the on-axis electron temperature of $T_e \approx 5$ keV is shown to drop down to $\approx 2$ keV. The impact of the sawtooth crash is very weakly present near the inversion radius. In this manuscript, we analyze the plasma during time windows when the profiles have fully recovered after a sawtooth crash (dashed lines in figure \ref{magnetics}.(b)).

After recovery of the profiles from a sawtooth crash, the trapped $^3$He population at MeV energies was observed to destabilize a range of Alfvén eigenmodes during the medium- and high-$P_\text{ICRH}$ phases. Both phenomena are often simultaneously observed in tokamak experiments \cite{nabais_nf_2010, gassner_pop_2012, calado_nf_2022}. Figure \ref{magnetics}.(a) shows the line-integrated, density fluctuation spectrogram from the far-infrared (FIR) interferometer in JET \cite{boboc_rsi_2010}. During the medium-$P_\text{ICRH}$ phase, fluctuations associated with Alfvén modes are observed around $300$ kHz. During the high-$P_\text{ICRH}$ phase, separate branches are observed to sweep down in frequency in the range $ 300 \rightarrow 280$ kHz, consistent with the expected temporal evolution of the central safety factor following a sawtooth crash \cite{saigusa_ppcf_1995, saigusa_ppcf_1998}. This behavior has already been observed in many major tokamaks, e.g., JT-60U \cite{kimura_physlettA_1995, saigusa_ppcf_1995, kimura_nf_1998, saigusa_ppcf_1998, kramer_prl_1999}, TFTR \cite{nazikian_prl_2003}, and JET \cite{sharapov_pop_2002, sandquist_pop_2007, calado_nf_2022}. The modes are spatially localized around the $q=1$ surface with very low magnetic shear ($\hat{s} \approx 0.1-0.2$), as suggested by Doppler-backscattering measurements, pressure-constrained EFIT (\citep{szepesi_ukaea_efit}, including fast-ion pressure), and ECE. 

The unstable Alfvén eigenmodes in figure \ref{magnetics}.(a) appear after saturation of the plasma profiles following a sawtooth crash (before which the modes are expected to be stable or marginally stable), and have toroidal mode numbers in the range $n=3-6$. The toroidicity-induced gap frequency associated with the toroidal Alfvén eigenmode (TAE \cite{cheng_annphys_1985}) for large-aspect-ratio tokamaks is $f_\text{TAE} \approx v_A / (4 \pi q R_0)$, where $R_0$ is the major radius of the flux-surface centre. The numerical values for the density, $q =1$ and $R_0\approx 2.9$ m lead to $f_\text{TAE} \approx 270$ kHz. This is consistent with the fluctuations seen in figure \ref{magnetics}.(a), which suggests that these modes live in the TAE gap, and could be identified as tornado TAEs \cite{saigusa_ppcf_1998}. Note that the plasma beta in this discharge is low ($\beta_e \lesssim 0.004$ for electrons), which is consistent with the conditions to excite shear Alfvén modes such as TAEs.

The dominant electron heating in this discharge increases the electron temperature at the inner radii from 2 keV (Ohmic) to 5 keV (high $P_\text{ICRH}$). The main-ion temperature in the deep core also increases (figure \ref{teti_time}.(a)) and approaches the electron temperature at medium and high $P_\text{ICRH}$. The measurements are made with main-ion charge-exchange recombination spectroscopy \cite{hawkes_rsi_2018} during $10$ ms neutral-beam-injection (NBI) blips (performed every second), which allow the temperature to be measured without affecting the overall plasma. The increase in the main-ion temperature shows that it is not clamped at a specific value, as in recent experiments of pure electron-cyclotron-resonance heating \cite{beurskens_nf_2021, beurskens_nf_2022}. This is a very positive result for fusion, since burning plasmas will have dominant electron heating from the slowing down of 3.5 MeV alpha particles, but will require high main-ion temperatures for fusion reactions to occur. The main ions here also exhibit an increase in the normalized temperature gradient $a/L_{Ti} = -\text{d} \log T_i/\text{d}\rho$ inside $\rho\approx 0.4$ ($a$ is the plasma minor radius), which starts outside the sawtooth inversion radius at $\rho \approx 0.25-0.3$. Additionally, detailed analysis of power balance using TRANSP shows that the ion heat diffusivity $\chi_i$ is significantly reduced, particularly in the inner core, as shown by figure \ref{teti_time}.(d) (the reduction in $\chi_i$ is accompanied by reduced ion heat losses, not shown). Note that the stabilization observed in the main-ion heat diffusivity takes place well outside the sawtooth inversion radius, which confirms that sawteeth are not the main mechanism affecting the temperature here. These findings support the thesis that fast-ion-driven modes improve the energy confinement of the thermal plasma. 

The increase in the main-ion temperature is a striking finding. The main ions experience negligible direct heating from ICRH, and their collisional coupling to the electrons decreases as the temperature increases. This is demonstrated in figure \ref{magnetics}.(b), where the electron-ion collision time $\tau_{ei}$ divided by the total energy-confinement time $\tau_E$ is shown to increase (as modeled by TRANSP). Linear Landau damping of the TAE on the main ions is also expected to be very weak due to low ion beta \cite{fu_cheng_physfluB_1992, connor_eps_1994}, as is nonlinear Landau damping \cite{bierwage_prl_2015, hahm_prl_1995, hahm_pst_2015, seo_nf_2021} (here $ \beta_i \lesssim 0.003$). This rules out the possibility of alpha channeling \cite{fisch_nf_1994, fisch_ppcf_1999}. Another possibility is the turbulent energy exchange between electrons and ions \cite{hinton_pop_2006, waltz_pop_2008, candy_pop_2013}. This is expected to be small in current devices \cite{waltz_pop_2008, candy_pop_2013}, and to cool the ions and heat the electrons in reactor-relevant conditions \cite{kato_arxiv_2024}, in contrast to the experimental observations presented here. This suggests that a different mechanism must explain the ion-temperature increase. The total energy-confinement factor for the L-mode is shown to be $H_{89,P} < 1$ in the low-$P_\text{ICRH}$ case, but increases to $H_{89,P} > 1$ with increasing $P_\text{ICRH}$ (figure \ref{magnetics}.(b)). The improvement of confinement and ion-temperature increase cannot be explained by previous mechanisms affecting the linear micro-instability \citep{citrin_ppcf_2023} (see appendix \ref{lin_cgyro} for a study of the effect of energetic $^3$He on the linear stabilty using the CGYRO code \cite{candy_jcp_2016} and the Pyrokinetics framework \cite{patel_pyrokinetics_2024}), and we conjecture that these observations are due to the suppression of turbulent transport by zero-frequency zonal modes pumped by Alfvén eigenmodes \cite{disiena_nf_2019, mazzi_natphys_2022, biancalani_ppcf_2021, biancalani_jpp_2023}. Below, we provide direct experimental evidence for the presence of such a mechanism. We stress that this situation is very different from NBI-heated discharges that use lower-energy fast ions, especially at low beta, where most of the heating is absorbed by the thermal ions, external torque drives flow-shear suppression of the turbulence, and Alfvén modes are more difficult to destabilize (although it is possible for NBI fast ions to marginally destabilize Alfvén modes, and numerical simulations have shown that these modes can nonlinearly interact with the turbulence \cite{citrin_ppcf_2023}, especially at high beta or with substantial populations of energetic ions). Similar experiments in JET with MeV-range ions have exhibited an improvement in the confinement of the thermal plasma \cite{nocente_nf_2020, kazakov_pop_2021, mazzi_natphys_2022}. 


To confirm that the Alfvén modes shown in figure \ref{magnetics} play a crucial role in the overall confinement, we performed turbulence measurements at $\rho \approx 0.25-0.35$ using the Doppler-backscattering system (DBS) in JET, which detects fluctuations of the electron density. In the Doppler-backscattering technique \citep{holzhauer_ppcf_1998, hirsch_rsi_2001, hennequin_nf_2006, hillesheim_nf_2015b, hirsch_ppcf_2004, hennequin_rsi_2004, hillesheim_prl_2016}, a beam of microwaves is launched into the core plasma until it encounters a cutoff surface. The detector receives backscattered radiation from turbulent fluctuations with characteristic wavenumber $k_\perp$ that are local to the vicinity of the cutoff \cite{valerian_ppcf_2022, ruizruiz_jpp_2024}, as predicted by beam-tracing numerical simulations of the microwave beam \cite{valerian_ppcf_2022}. We normalize the scattered turbulence wavenumber $k_\perp$ by the local ion sound gyroradius $\rho_s = c_s/\Omega_D$, where $c_s = \sqrt{T_e/m_D}$ is the sound speed, $m_D$ is the deuterium mass, $\Omega_D = q_D B_0/(m_D c)$ its gyro-frequency, $q_D$ its charge, and $c$ the speed of light. The DBS measurements presented here are sensitive to density fluctuations at wavenumbers $k_\perp\rho_s \approx 2-4$. Dedicated analysis of the scattering trajectory using the beam-tracing code Scotty \cite{valerian_ppcf_2022, ruizruiz_jpp_2024} shows that $k_\perp$ has a toroidal component that overlaps with $n=3-6$ of the unstable Alfvén modes in figure \ref{magnetics}.(a) ($k_\perp$ is in fact mostly in the direction normal to the flux surface, while $n=3-6$ are well within the range given by the diagnostic's resolution). This suggests that DBS measurements can be sensitive to Alfvén modes in these conditions.

\begin{figure}
\begin{center}
	\includegraphics[height=6cm]{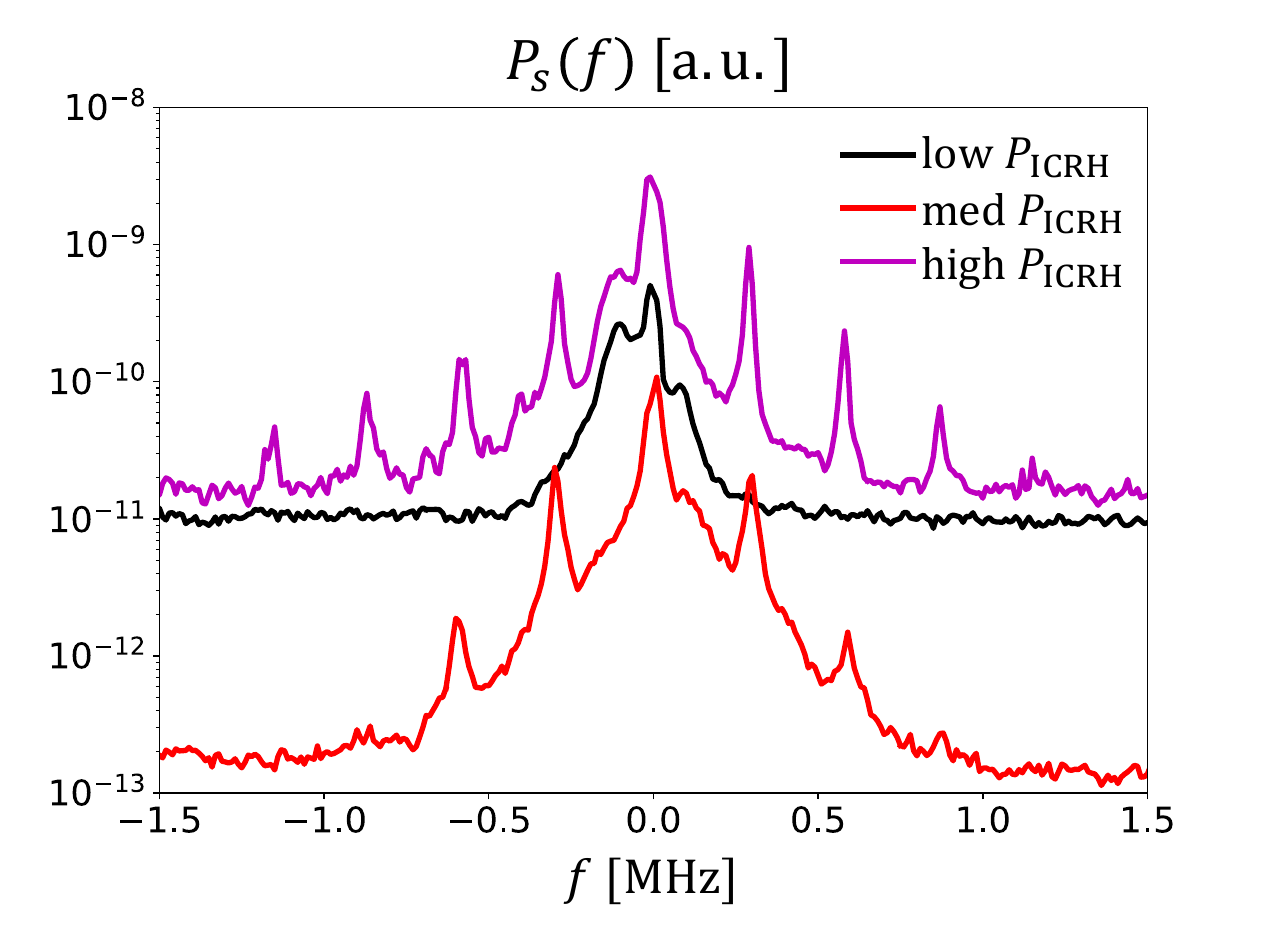}
\end{center}
\caption{Power spectra $P_s(f)$ at low, medium, and high $P_\text{ICRH}$. At low $P_\text{ICRH}$, the spectrum is broadband, characteristic of drift-wave turbulence. At medium $P_\text{ICRH}$, note spectral peaks at $\pm f_\text{TAE} $ and $\pm 2f_\text{TAE}$. At high $P_\text{ICRH}$, $P_s(f)$ displays spectral peaks at $f_\text{TAE}$ and its multiples (up to four). }   
\label{dbs_low}
\end{figure}

\begin{figure}
\begin{center}
	\includegraphics[height=9cm]{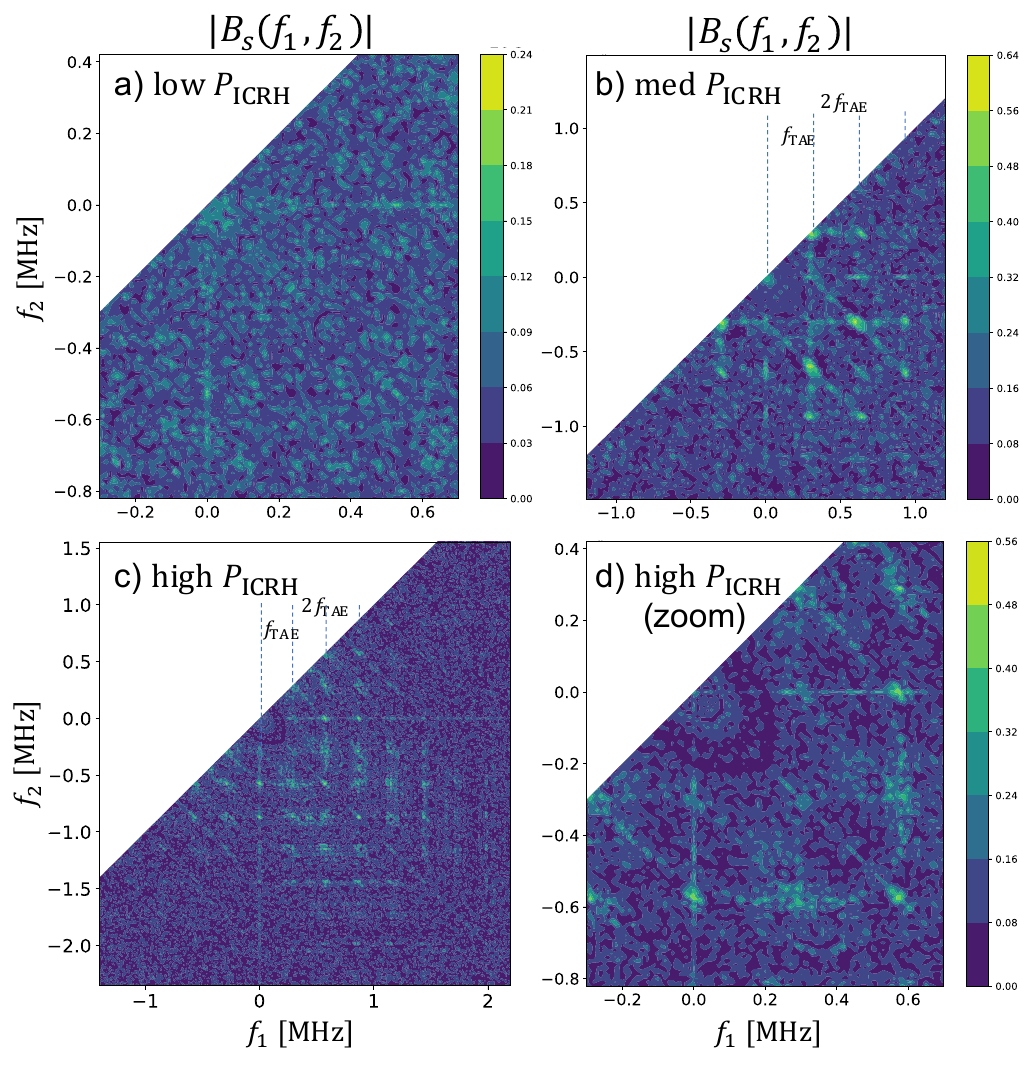}
\end{center}
\caption{Bicoherence spectra at low (a), medium (b), and high $P_\text{ICRH}$ (c)-(d). At low $P_\text{ICRH}$, we observe broadband phase-matched relationship in the range $f_1, f_2 < 100$ kHz. At medium $P_\text{ICRH}$, the bicoherence displays phase-matched relationships for multiples of $f_\text{TAE}$ (vertical dashed lines), the strongest of which at $\pm f_\text{TAE}$ and $\pm2f_\text{TAE}$. A weaker phase-matched relationship is observed between fluctuations at multiples of $f_\text{TAE}$ and a zero-frequency fluctuation. At high $P_\text{ICRH}$, the bicoherence displays phase-matched relationships for multiples of $f_\text{TAE}$. The strongest phase matching is for frequency pairs involving a zero-frequency fluctuation: $(f_1, f_2) = \{(\pm 2f_\text{TAE},0),(\pm3f_\text{TAE}, 0)\}$. Panel (d) zooms into panel (c).}   
\label{dbs_med}
\end{figure}

Figures \ref{dbs_low} and \ref{dbs_med} show the power spectrum $P_s(f)$ and the bicoherence $B_s(f_1, f_2)$ of the electron-density fluctuations for the low-, medium- and high-$P_\text{ICRH}$ phases. The measurements are taken at specific times that are shown by the colored vertical dashed lines in figure \ref{magnetics}.(b). The backscattered power is calculated from the Fourier-transformed backscattered complex signal amplitude $\hat{A}_{j}$ as $P_s(f) = \langle |\hat{A}_{j}|^2 \rangle_T$, where the $\langle . \rangle_T$ denotes an ensemble average and the $f=0$ component of $\hat{A}_{j}$ is removed ($j$ denotes the sample realization). The period of each time window is $T=0.05-0.1$ ms, which is much longer than the turbulence correlation time, which scales as $a/c_s \sim 10^{-6}$ s. The bicoherence \cite{kim_powers_physflu_1978} associated with the ensemble of realization signals $\{ A_j \}$ is 

\begin{equation}
    \begin{alignedat}{4}
    & B_s(f_1, f_2) = \frac{\langle \hat{A}_{j}(f_1)\hat{A}_{j}(f_2) \hat{A}_{j}(f_1+f_2)^* \rangle_T}{\langle |\hat{A}_{j}(f_1)\hat{A}_{j}(f_2)|^2 \rangle_T^\frac{1}{2} \langle |\hat{A}_{j}(f_1+f_2)|^2 \rangle_T^\frac{1}{2} } .
    \end{alignedat}
    \label{bicoh_eq}
\end{equation}
The modulus of the bicoherence $|B_s(f_1, f_2)|$ in equation \eqref{bicoh_eq} is high when the frequency $f_1$ has a phase-matched relationship with the frequency $f_2$ to generate a third frequency at $f_1 \pm f_2$. The ensemble averaging $\langle . \rangle_T$ is critical to eliminate independent modes at the frequencies $f_1 \pm f_2$. The bicoherence is a measure of three-wave nonlinear coupling between fluctuations of frequency $f_1$ and $f_2$.

Figure \ref{dbs_low} shows the power spectrum $P_s(f)$ (in arbitrary units [a.u.]) at low, medium, and high $P_\text{ICRH}$. At low $P_\text{ICRH}$, the DBS beam reaches the cutoff surface at $\rho = 0.247$ (where $q\approx 1$), corresponding to the scattered wavenumber $k_\perp \rho_s = 2.58$. The power spectrum $P_s(f)$ is broadband for fluctuations in the range $f \lesssim 200 $ kHz characteristic of drift-wave turbulence \citep{liewer_nf_1985, tynan_ppcf_2009}, and particularly, turbulence driven by the ion-temperature gradient mode (ITG) (note that non-broadband, drift-wave-like turbulence associated with other linear micro-instabilities is reported in the literature \cite{arnichand_ppcf_2016, kotschenreuther_nf_2019}, but appendix \ref{lin_cgyro} shows that those are not relevant here). A peak at zero frequency is observed, consistent with many previous DBS measurements. At medium $P_\text{ICRH}$, the DBS beam reaches the cutoff surface at $\rho = 0.355$ ($q\approx 1$), corresponding to the scattered wavenumber $k_\perp \rho_s = 2.07$. The power spectrum $P_s(f)$ is broader than in the low-$P_\text{ICRH}$ phase, featuring fluctuations in the range $f \lesssim 800 $ kHz, as well as spectral peaks at the local Alfvén gap frequency $f_\text{TAE} \approx 290 $ kHz and a harmonic at $2f_\text{TAE}$. The spectral peaks appear for positive and negative frequencies, which could suggest the presence of bidirectional Alfvénic activity. Careful analysis of the DBS microwave beam, pressure-constrained EFIT, and the $q=1$ constraint from ECE shows that the modes observed in figure \ref{dbs_low} are well localized around $q=1$. This is consistent with previous observations of core-localized TAEs in TFTR \cite{fredrickson_nf_1995, fu_pop_1996, nazikian_pop_1998} and tornado modes in JT-60U \cite{saigusa_ppcf_1998}. Such modes were predicted to exist at low shear \cite{berk_pop_1995}, consistent with the low shear value in this JET experiment. At high $P_\text{ICRH}$, the DBS beam reaches the cutoff surface at $\rho = 0.251$ ($q \approx 0.92$) for $k_\perp \rho_s = 3.78$. The power spectrum is broadband for fluctuations in the range $f \lesssim 600 $ kHz, but also displays spectral peaks at the local Alfvén gap frequency $f_\text{TAE} \approx 290 $ kHz and its multiples, which now extend to four (the difference in the background noise level of the curves in figure \ref{dbs_low} originates from the difference in the absolute power launched from each of the corresponding DBS channels). 

Figure \ref{dbs_med} shows the modulus of the bicoherence $|B_s(f_1, f_2)|$ for low, medium and high $P_\text{ICRH}$. At low $P_\text{ICRH}$ in \ref{dbs_med}.(a), the bicoherence is calculated for $255$ windows of period $T=0.1$ ms that are half-overlapping. Its modulus displays broadband phase-matched relationships between frequencies in the range $f_1, f_2 < 100$ kHz, as expected from nonlinear interactions between turbulent eddies characteristic of drift-wave turbulence. The amplitude, however, remains low at $\approx 0.2$, and this suggests that nonlinear interactions between low-frequency drift-wave fluctuations are broadband and modest in the low-$P_\text{ICRH}$ phase. At medium $P_\text{ICRH}$ in \ref{dbs_med}.(b), the bicoherence is calculated for a total of $63$ half-overlapping windows of period $T=0.05$ ms. It displays broadband phase-matched relationships for multiples of the local Alfvén gap frequency $f_\text{TAE}$, the strongest of which are $(f_1, f_2) = \{(\pm f_\text{TAE}, \pm f_\text{TAE})\}$ (amplitude $\approx 0.6$), which correspond to co-propagating TAEs that couple to generate a perturbation of frequency $\pm 2f_\text{TAE}$, as well as $(f_1, f_2) = \{(2f_\text{TAE}, - f_\text{TAE}), (- f_\text{TAE}, 2f_\text{TAE}) \}$, which correspond to perturbations of frequency $2f_\text{TAE}$ coupling to a counter-propagating perturbation at frequency $f_\text{TAE}$ that generates a co-propagating perturbation at $f_\text{TAE}$. The bicoherence also clearly exhibits phase-matched relationships between perturbations at multiples of $f_\text{TAE}$ and a low-frequency fluctuation, which we call the \emph{zero-frequency fluctuation} in the rest of the manuscript. To our knowledge, this is the first experimental demonstration in a magnetically confined plasma that Alfvén modes living inside the fundamental toroidicity-induced frequency gap (and higher-frequency gaps) beat with counter-propagating perturbations to generate a zero-frequency fluctuation. The zero-frequency fluctuations observed here can naturally be interpreted as zonal modes driven by Alfvén eigenmodes, which have been predicted analytically \cite{chen_prl_2012, qiu_pop_2016,qiu_nf_2017, qiu_revmodphys_2023} and numerically \cite{spong_pop_1994, todo_nf_2012, zhang_pst_2013, biancalani_iaea_2016, mazzi_natphys_2022, biancalani_ppcf_2021, biancalani_jpp_2023}. Such zonal modes can suppress the turbulence driven by ITG, which would explain the increase of the ion temperature and its gradient (figure \ref{teti_time}). At high $P_\text{ICRH}$ in \ref{dbs_med}.(c)-(d), the bicoherence is calculated for a total of $127$ half-overlapping windows of period $T=0.1$ ms. Similarly to \ref{dbs_med}.(b), the bicoherence displays broadband phase-matched relationships for multiples of $f_\text{TAE}$. Importantly, the highest-amplitude phase matching is now for the frequency pairs $(f_1, f_2) = \{ (2f_\text{TAE}, 0), (2f_\text{TAE}, -2f_\text{TAE}), (3f_\text{TAE}, 0), (3f_\text{TAE}, -3f_\text{TAE}) \} $ (amplitude $\approx 0.5$), i.e., the pairs that involve a zero-frequency fluctuation. This is different from the medium-$P_\text{ICRH}$ phase, in which the phase-matched relationships involving the zero-frequency fluctuation were subdominant to other mode-mode interactions. Interestingly, the modes at the fundamental Alfvén gap frequency $f_\text{TAE}$ display a weaker phase-matched relationship with the zero-frequency fluctuation. Higher multiples of the Alfvén frequency show weaker phase-matching, resulting in a peculiar grid structure. 

Previous studies have observed bidirectional Alfvén modes in a tokamak \cite{saigusa_ppcf_1998,sandquist_pop_2007}, but the nonlinear coupling with a zero-frequency zonal mode was not shown. TAEs and energetic-particle modes were also reported to interact nonlinearly with each other in \cite{crocker_prl_2006}, but they were not shown to couple nonlinearly to a zero-frequency zonal mode. In a recent publication \cite{garcia_natcomm_2024}, numerical simulations showed that TAEs could generate a strong zonal flow and stabilize the turbulence, leading to an enhanced-confinement DT plasma in JET. This is a promising result for extrapolation to burning plasmas, but it lacked the experimental verification of a zonal flow, which we provide here.

Figures \ref{dbs_med}.(b)-(d) confirm that Alfvén modes living inside the toroidicity-induced Alfvén-frequency gap (and higher-frequency gaps) beat with counter-propagating Alfvén modes to generate a zero-frequency fluctuation that is consistent with a zonal mode. This behavior is \emph{correlated} with an increase in the total energy-confinement factor ($H_{89,P} > 1$), the main-ion temperature and its gradient (despite negligible direct ion heating), and a reduction in the main-ion heat diffusivity. This shows the strong beneficial effect that Alfvén eigenmodes can have on the confinement of the thermal plasma. This is to be constrasted with the stable-Alfvén-mode case (low $P_\text{ICRH}$), which exhibited low ion temperature and temperature gradient, and $H_{89,P} < 1$. This is the first experimental confirmation in a magnetically confined plasma of a zero-frequency fluctuation that is pumped by Alfvén modes, and is correlated with hotter-than-expected main ions. It suggests that zonal modes driven by Alfvén modes can suppress the turbulence and lead to an overall improvement of energy confinement.

Admittedly, the presence of unstable Alfvén eigenmodes is not always correlated with improved confinement. It is also responsible for strong energetic-particle transport, which is not studied here, and is highly detrimental to energy confinement. The balance between this and the improved confinement for the thermal plasma observed here is still an open problem. It seems plausible that if this balance is quantitatively understood, unstable Alfvén modes could be tailored to improve the overall confinement in the burning plasmas expected to be achieved in the imminent future.

This work has been supported by the Engineering and Physical Sciences Research Council (EPSRC) [EP/W026341/1]. The work of M.B. and A.A.S. was supported in part by the EPSRC grant EP/R034737/1. The work of A.A.S. was also supported in part by the STFC grant STW000903/1 and by the Simons Foundation via a Simons Investigator Award. V. H. Hall-Chen was partly funded by an A*STAR Green Seed Fund, C231718014, and a YIRG, M23M7c0127. This work was supported by the U.S. Department of Energy under contract number DE-AC02-09CH11466. The United States Government retains a non-exclusive, paid-up, irrevocable, world-wide license to publish or reproduce the published form of this manuscript, or allow others to do so, for United States Government purposes. This work has been carried out within the framework of the EUROfusion Consortium, funded by the European Union via the Euratom Research and Training Programme (Grant Agreement No 101052200 — EUROfusion). Views and opinions expressed are however those of the author(s) only and do not necessarily reflect those of the European Union or the European Commission. Neither the European Union nor the European Commission can be held responsible for them.


\appendix

\section{Linear gyrokinetic calculations}
\label{lin_cgyro}

In this appendix, we confirm that the increase in the ion temperature and its gradient, the improvement of confinement, and the reduction in the ion heat diffusivity observed in the JET plasma presented in this letter cannot be explained by previous mechanisms of \emph{direct} stabilization of turbulence by energetic particles \citep{citrin_ppcf_2023}: dilution of the main plasma is negligible for the density of $n_{^3\text{He}}/n_e \approx 0.2-0.3 \%$; stabilization by beta-related mechanisms is also negligible due to the low electron beta $\beta_e \lesssim 0.004$; and the fast-ion pressure and its gradient have a negligible effect on the equilibrium. The mechanisms of linear stabilization described in \citep{disiena_nf_2018, wilkie_nf_2018} are not relevant either, since those effects are only significant at energetic-particle energies $E/T_e \approx 10$, in contrast to $E/T_e \gtrsim 10^2-10^3$ in the deep core of the plasma that we have studied. These \emph{linear}, direct effects of the energetic particles have a negligible effect on the linear microinstability driven by the ion-temperature gradient (ITG), as we show in this appendix. 

\begin{figure}
\begin{center}
	\includegraphics[height=4.5cm]{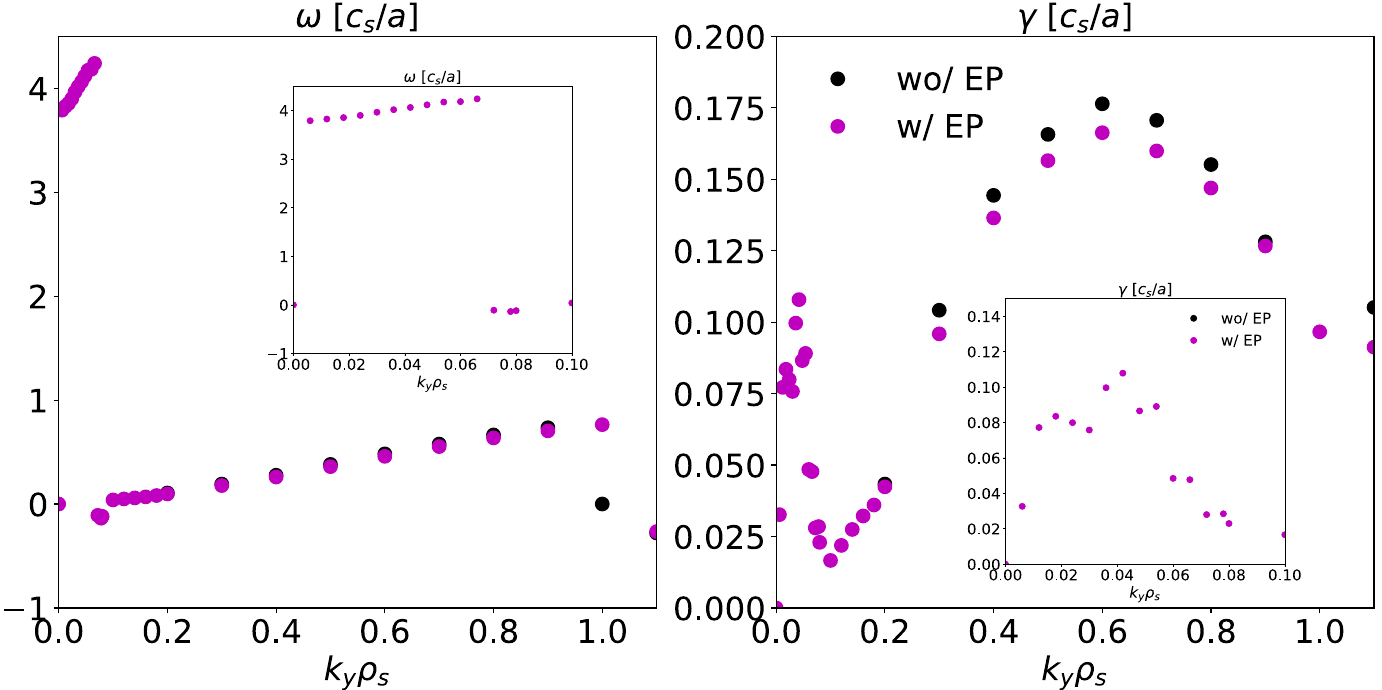}
\end{center}
\caption{Left panel: real-frequency $\omega$ vs. poloidal wavenumber $k_y\rho_s$ of the fastest-growing mode. Right panel: corresponding linear growth rate $\gamma$. Insets zoom in on the low-$k_y$ part of the spectrum. The case with a population of fast $^3$He ions is shown as magenta points, the case without as black ones.}   
\label{cgyro_lin_w_wo_fi}
\end{figure}

In order to prove that linear mechanisms are negligible in this JET configuration, we carry out linear micro-stability calculation of the most-unstable mode using the gyrokinetic code CGYRO \cite{candy_jcp_2016} and the Pyrokinetics framework \cite{patel_pyrokinetics_2024}. We extract the plasma parameters from the high-$P_\text{ICRH}$ case at the radial location $\rho\approx0.3$, which overlaps with the experimental measurements of the zero-frequency fluctuations. Figure \ref{cgyro_lin_w_wo_fi} shows the linear spectrum of the real frequency $\omega$ and growth rate $\gamma$ normalized by $c_s/a$, where $c_s = \sqrt{T_e/m_D}$ is the ion sound speed, $a$ is the minor radius, $m_D$ is the deuterium mass and $T_e$ is the local electron temperature. Two types of numerical calculations are performed: with a population of fast $^3$He ions, denoted \enquote{w/ EP} modelled by a Maxwellian of temperature equal to the mean energy of the fast particles, as provided by TRANSP-TORIC; and without the fast population of $^3$He, denoted \enquote{wo/ EP} (black). Both simulation types resolve thermal deuterium, hydrogen and electrons as gyrokinetic species. In the simulations \enquote{wo/ EP}, the value of the radial gradient of beta is scaled to maintain self-consistency of the species gradients with the magnetic equilibrium, and the density of the hydrogen is scaled to satisfy the quasi-neutrality condition (this modification is $< 1\%$, due to the low density of $^3$He). 

The linear calculation without fast $^3$He shows a dominant instability whose growth rate peaks at $k_y\rho_s \approx 0.6$, driven in the ion-diamagnetic drift direction $\omega>0$, consistent with the ITG mode. The simulations with fast $^3$He exhibit a very similar mode, but reduced by a factor of $< 5\%$. This reduction is consistent with dilution, the effect of the gradient of beta on the equilibrium, and the linear kinetic effect discussed in \citep{disiena_nf_2018, wilkie_nf_2018}. Additionally, at $k_y\rho_s \approx 0.01-0.05$ (see inset), a new destabilized mode appears, which is consistent with a TAE. Using the experimental plasma parameters, one can calculate the toroidal mode number $n$ that corresponds to the simulated $k_y\rho_s$. This leads to the prediction of unstable modes in the range of $n \approx 1-11$, consistent with the experimental observations of TAEs in figure 1 of the main manuscript (note that the simulation was run with the temperature gradient of $^3$He $a/L_{T_{^3\text{He}}} = 15.7$ provided by TRANSP-TORIC, which is unrealistically large; $a/L_{T_{^3\text{He}}}$ is expected to be closer to marginal stability, which would reduce the growth rate and the range of toroidal mode numbers driven unstable). The real frequency of these modes is $\omega \approx 4c_s/a$. The dimensional value of the frequency using experimental parameters is $f \approx 260-290$ kHz, which is once more consistent with the experimental frequency range of the modes presented in figure 1 of the main manuscript.

The results of this appendix confirm that dilution, the effect of the gradient of beta on the equilibrium, and the linear kinetic effects of \cite{disiena_nf_2018, wilkie_nf_2018} are negligible here: combined, they give rise to a $< 5\%$ reduction of the linear growth rate of the fastest growing micro-instability. Having ruled out these possibilities leaves us with the effect of the zonal flow pumped by Alfvén modes as the only mechanism able to explain the ion-temperature increase and the improvement of confinement in this JET plasma. The experimental measurements of the zero-frequency fluctuation support this conclusion. 

\bibliographystyle{apsrev4-2}  
\bibliography{mybib_arxiv_v2}   

\end{document}